\documentclass[prd,twocolumn]{revtex4}
\usepackage{graphicx, epsfig}
\usepackage{color}
\usepackage{mathrsfs}
\usepackage{amsmath}

\newcommand{\be}{\begin{equation}}
\newcommand{\ee}{\end{equation}}
\newcommand{\bea}{\begin{eqnarray}}
\newcommand{\eea}{\end{eqnarray}}

\newcommand{\gapp}{\mathrel{\raise.3ex\hbox{$>$}\mkern-14mu
              \lower0.6ex\hbox{$\sim$}}}
\newcommand{\lapp}{\mathrel{\raise.3ex\hbox{$<$}\mkern-14mu
              \lower0.6ex\hbox{$\sim$}}}

\begin{document}

\title{Time dependent fluctuations and particle production in cosmological de Sitter and anti-de Sitter spaces}
\author{Eric Greenwood}
\affiliation{CERCA, Department of Physics, Case Western Reserve University, Cleveland, OH 44106-7079}
\author{De Chang Dai}
\author{Dejan Stojkovic}
\affiliation{HEPCOS, Department of Physics,
SUNY at Buffalo, Buffalo, NY 14260-1500}
\begin{abstract}
We study the evolution of time-dependent fluctuations and particle production in an expanding dS and contracting AdS universe. Using the functional Schrodinger formalism we are able to probe the time dependent regime which is out of the reach of the standard approximations like the Bogolyubov method. In both cases, the evolution of fluctuations is governed by the harmonic oscillator equation with time dependent frequency. In the case of an expanding dS universe we explicitly show that the frequency of fluctuations produced at a certain moment diminish in time, while the distribution of the created particles quickly approaches the thermal radiation of the dS space.  In the case of a contracting AdS universe we show that the frequency of fluctuations produced at a certain moment grow in time. Nominally, the temperature of radiation diverges as the Big Crunch is approaching, however, increasing oscillations of the spectrum make the temperature poorly defined, which is in agreement with the fact that AdS space does not have an event horizon which would cause thermal radiation. Unlimited growth of fluctuations indicates that an eventual tunneling into AdS vacuum would have catastrophic consequences for our universe.

\end{abstract}


\maketitle

\section{Introduction}

Space-times with the constant vacuum energy density play very important role in modern cosmology. It is believed that the primordial inflation was driven by the cosmological constant or a scalar field vacuum energy density of positive sign. As the recent observational data indicate, our late time universe may again be dominated by the (positive) vacuum energy density which drives the accelerated expansion of the universe. Both primordial and late time phases of acceleration can be effectively described by de Sitter (dS) space-time. On the other hand, as indicated by string theory and other generic considerations \cite{Stojkovic:2007dw}, it may happen that the true vacuum energy density of our universe is negative, which is described by anti-de Sitter (AdS) space. In that case, our universe will in the future undergo a phase transition from dS to AdS space.

dS space describes an expanding universe. Time dependent gravitation field usually leads to particle production. Particles that propagate in the background of a time dependent metric of an expanding universe get excited and radiation is produced. For example, this is the mechanism behind particle production in FRW universe. In addition to this, dS space contains an event horizon (called dS horizon, not to be confused with the particle horizon). Using an analogy with the event horizons in black hole space-times, one could expect that the produced radiation of dS space is thermal with the constant temperature which is proportional to the value of the cosmological constant $\lambda$. Indeed, such a result may be obtained using standard approximations like the Bogolyubov method or tunneling formalism. However, time-dependent evolution of radiation is beyond the reach of these standard approximations. The main goal of this paper is to study these time-dependent effects. We will use the recently developed functional Schrodinger formalism \cite{FSF} to study a scalar field propagating in the background of an expanding dS space. We will show that the evolution of fluctuations (excitations) of the scalar field is governed by the harmonic oscillator equation with the time dependent frequency. We will solve the equations of motion to find the time-dependent wave function for the system. If we set the wave function of the system to be in the ground state at a certain moment of time, the particle content at some later time will be given by the wave function overlap between the initial and final state. This wave function overlap will give us the time-dependent occupation number of produced particles. In cosmology one is very often interested in the power spectrum, i.e. in power contained in a mode of a certain frequency, but this quantity is very closely related to the occupation number of particles in a given mode \cite{Mahajan:2007qg}. We will show that when a mode of ceratin frequency is created it does not automatically follow a thermal distribution with the temperature proportional to the value of the Hubble parameter (which is at late times proportional to the cosmological constant $\lambda$). Modes whose frequencies are much larger or much smaller than the Hubble parameter do not follow a thermal distribution, while those of the order of the Hubble parameter do. As the high frequency modes get stretched by expansion, and their frequencies become comparable to the Hubble parameter, they become thermal.

We will then apply the same technique to AdS space.
In the cosmological context, AdS space may be used to describe a collapsing universe. Unlike dS space, AdS space does not contain an event horizon, so we do not expect thermal radiation with the constant temperature. However, a time-dependent metric may again lead to particle production. The fate of the universe which tunnels into the true vacuum which is AdS was studied in \cite{Coleman:1980aw,Abbott:1985kr}. There, it was argued that after a short period of expansion, the universe starts recollapsing and ends up forming a black hole. Here, we study the time-evolution of the scalar field propagating in the background of AdS space and our findings support these arguments. Fluctuations of the scalar field produce more and more modes of high frequency at late times. The best fit temperature of the produced particles diverges at the time of the Big Crunch. However, towards the Big Crunch, oscillations of the spectrum grow in amplitude making the temperature less and less defined.

\section{Friedman-Robertson-Walker space-time}

\label{FRW}

To setup the formalism, we will first consider the standard Friedman-Robertson-Walker (FRW) space-time.
The action for the scalar field propagating in the curved space-time background given by the metric tensor $g_{\mu\nu}$ is
\be
  S=\int d^4x\sqrt{-g}\frac{1}{2}g^{\mu\nu}\partial_{\mu}\Phi\partial_{\nu}\Phi.
  \label{action}
\ee
Decomposing the (spherically symmetric) scalar field into a complete set of real basis function denoted by $\{f_k(r)\}$
\be
  \Phi=\sum_ka_k(t)f_k(r)
  \label{mode}
\ee
we want to find a complete set of independent eigenmodes $\{b_k\}$ (which are linear combinations of the original modes $\{a_k\}$)  for which the Hamiltonian is a simple sum of the independent modes. The total wavefunction then factorizes and can be found by solving a time-dependent Schr\"odinger equation of just one variable.

The metric for the spherically symmetric FRW space-time is
\be
  ds^2=-dt^2+R(t)^2\left(\frac{dr^2}{1-kr^2}+r^2d\Omega^2\right)
  \label{metric}
\ee
where
\be
  d\Omega^2=d\theta^2+\sin^2\theta d\phi^2 ,
\ee
$R(t)$ is the scale factor, while $k =-1,0,1$. Using Eq.~(\ref{metric}) and Eq.~(\ref{action}) we can write
\be
\label{a1}
  S=2\pi\int dt\int dr r^2 \frac{R^3(t)}{\sqrt{1-kr^2}}\left[-(\partial_{\tau}\Phi)^2+\frac{1-kr^2}{R^2(t)}(\partial_r\Phi)^2\right].
\ee

Using the mode expansion in Eq.~(\ref{mode}), we can rewrite the action as
\be
  S=\int dt\left[-\frac{R^3(t)}{2}\dot{a}(t)_k{\bf M}_{kk'}\dot{a}(t)_{k'}+\frac{R(t)}{2}a(t)_k{\bf N}_{kk'}a(t)_{k'}\right]
\ee
where $\dot{a}(t)=da(t)/dt$, and ${\bf M}$ and ${\bf N}$ are matrices that are independent of $R(t)$ and are given by
\bea
  {\bf M}_{kk'}&=&4\pi\int drr^2\frac{1}{\sqrt{1-kr^2}}f_kf_{k'},\label{M}\\
  {\bf N}_{kk'}&=&4\pi\int drr^2\sqrt{1-kr^2}f'_kf'_{k'}.
  \label{N}
\eea
For our procedure we will use the fact that the matrices ${\bf M}$ and ${\bf N}$ are symmetric, while their exact form will not be needed. From the action Eq.~(\ref{a1}) we can find the Hamiltonian, H, and write the Schr\"odinger equation
\be
H(a_k,t) \psi(a_k,t)=i\frac{\partial\psi (a_k,t) }{\partial t} .
\ee
The wavefunction $\psi(a_k,t)$ is actually the wavefunctional since $a_k(t)$ is a function itself. Therefore, this procedure is the functional Schr\"odinger formalism. Defining $\Pi_k$, the momentum operator conjugate to $a(t)_k$ as
\be
  \Pi_k=-i\frac{\partial}{\partial a(t)_k}
\ee
the  Schr\"odinger equation can be written as
\be \label{nd}
  \left[-\frac{1}{2R^3(t)}\Pi_k({\bf M}^{-1})_{kk'}\Pi_{k'}+\frac{R(t)}{2}a(t)_k{\bf N}_{kk'}a(t)_{k'}\right]\psi=i\frac{\partial\psi}{\partial t}
\ee
The principal axis transformation guaranties that ${\bf M}$ and ${\bf N}$ can be simultaneously diagonalized. Therefore, the Schr\"odinger equation Eq.~(\ref{nd}) decouples into an infinite set of decoupled equations. The single eigenmode Schr\"odinger equation is
\be
  \left[-\frac{1}{2m}\frac{\partial^2}{\partial b^2}+\frac{K}{2}R^4(t)b^2\right]\psi(b,t)=iR^3(t)\frac{\partial\psi(b,t)}{\partial t}
  \label{schrod}
\ee
where $m$ and $K$ denote eigenvalues of ${\bf M}$ and ${\bf N}$, and $b$ is the eigenmode (which is a linear combination of the original modes $a_{k}$).

Re-writing Eq.~(\ref{schrod}) in the standard harmonic oscillator form (with the time dependent frequency) we obtain
\be
  \left[-\frac{1}{2m}\frac{\partial^2}{\partial b^2}+\frac{m}{2}\omega^2(\eta)b^2\right]\psi(b,\eta)=i\frac{\partial\psi(b,\eta)}{\partial\eta}
  \label{Schrod}
\ee
where
\be
  \omega^2(\eta)=\frac{K}{m}R^4(t)\equiv\omega^2_0R^4(t)
  \label{omega}
\ee
and
\be
  \eta=\int_0^{t}\frac{dt'}{R^3(t')}.
  \label{eta}
\ee
Obviously, $\omega_0=\sqrt{\frac{K}{m}}$ is the initial frequency of the mode at the time of its creation.
The exact solution to Eq.~(\ref{Schrod}) is given by \cite{Pedrosa}
\be
  \psi(b,\eta)=e^{i\alpha(\eta)}\left(\frac{m}{\pi\rho^2}\right)^{1/4}\exp\left[\frac{im}{2}\left(\frac{\rho_{\eta}}{\rho}+\frac{i}{\rho^2}\right)b^2\right]
  \label{psi}
\ee
where $\rho_{\eta}=d\rho/d\eta$ and $\rho$ is given by the real solution of the ordinary (though non-linear) differential equation
\be
  \rho_{\eta\eta}+\omega^2(\eta)\rho=\frac{1}{\rho^3}
  \label{rho}
\ee
with initial conditions
\be
  \rho(0)=\frac{1}{\sqrt{\omega_0}}, \hspace{2mm} \rho_{\eta}(0)=0.
  \label{IC}
\ee
The phase $\alpha$ is given by
\be
  \alpha(\eta)=-\frac{1}{2}\int_0^{\eta}\frac{d\eta'}{\rho^2(\eta')}.
  \label{alpha}
\ee

Consider an observer with detectors that are designed to register particles of different frequencies for the free field $\Phi$. Such an observer will interpret the wavefunction of a given mode $b$ at some later time in terms of simple harmonic oscillator states, $\{\varphi_n\}$, at the final frequency, $\bar{\omega}$. The initial ($t=0$) vacuum state for the modes is the simple harmonic oscillator ground state
\be
\varphi (b) = \left(\frac{m\omega_0}{\pi} \right)^{1/4} e^{-m\omega_0 b^2/2} \, .
\ee
The number of quanta in eigenmode $b$ can be evaluated by decomposing Eq.~(\ref{psi}) in terms of the states and evaluating the occupation number of that mode. Writing the wavefunction for a given mode in terms of simple harmonic oscillator basis at $t=0$ is given by
\be
  \psi(b,t)=\sum_n c_n(t)\varphi_n (b)
  \label{psi_ex}
\ee
where
\be
  c_n=\int db\varphi_n^*(b)\psi(b,t)
  \label{c_n}
\ee
which is an overlap of a Gaussian with the simple harmonic
oscillator basis functions. The occupation number at eigenfrequency $\bar{\omega}$ (which is the frequency at the final time of the measurement corresponding to $t_f$), is given by
\be
  N(t,\bar{\omega})=\sum_n n \left| c_n\right|^2.
  \label{occ_num}
\ee

The occupation number in the eigenmode $b$ is then given by (see Appendix
\ref{Number})
\be
  N(t,\bar{\omega})=\frac{\bar{\omega}\rho^2}{\sqrt{2}}\left[\left(1-\frac{1}{\bar{\omega}\rho^2}\right)^2+
  \left(\frac{\rho_{\eta}}{\bar{\omega}\rho}\right)^2\right] .
  \label{Occ_Num}
\ee
This occupation number represents the cumulative number of particles produced by the time $t_f$.
To extract the temperature corresponding to the occupation number distribution, we can compare Eq.~(\ref{Occ_Num}) with the Planck distribution
\be
  N_P(\omega)=\frac{1}{e^{\beta\omega}-1}
  \label{N_P}
\ee
where $\beta$ is the inverse temperature.

\section{Particle production in De Sitter space-time}

We now consider particle production in the background of dS space-time. The metric for dS space in static coordinates can be written as
\be
  ds^2=-\left( 1- \frac{r^2}{l^2} \right) dt^2+ \left( 1- \frac{r^2}{l^2} \right)^{-1} dr^2 + r^2d\Omega^2.
  \label{sc}
\ee
The constant $l$ is called dS radius.
Obviously, at $r=l$ the metric has an event horizon. This horizon is not to be confused with a particle horizon. The particle horizon represents the largest comoving distance from which light could have reached us by now, while the event horizon is the largest comoving distance from which light emitted now can ever reach the observer at any time in the future. Since black hole solutions with an event horizon are known to produce thermal radiation, it is expected that dS space can produce thermal radiation as well (for some counterarguments see \cite{Sekiwa:2008gk,Volovik:2008ww}).

We can also write the metric for dS space in non-static coordinates.
For this purpose, we set $k=0$ in Eq.~(\ref{metric}). The solution to the Einstein's equations with a positive cosmological constant $\lambda \sim 1/l$ give the expansion factor $R(t)\sim\exp(Ht)$, with $H$ (the Hubble parameter) given late times by
\be
  H\equiv\sqrt{\lambda/3}.
  \label{H}
\ee
Therefore, the time dependent dS metric is given by
\be
  ds^2=-dt^2+R^2(t)\left(dr^2+r^2d\Omega^2\right),
  \label{de_sit}
\ee
with $R(t)\sim\exp(Ht)$.
From here, we proceed like in Section \ref{FRW}.
Since $k=0$, the matrices ${\bf M}$ and ${\bf N}$, Eqs. (\ref{M}) and (\ref{N}) are now
\bea
  {\bf M}&=&4\pi\int drr^2f_kf_{k'},\label{de M}\\
  {\bf N}&=&4\pi\int drr^2f'_kf'_{k'}.\label{de N}
\eea
The occupation number is given again by Eq.~(\ref{Occ_Num}).

\begin{figure}[htbp]
\includegraphics{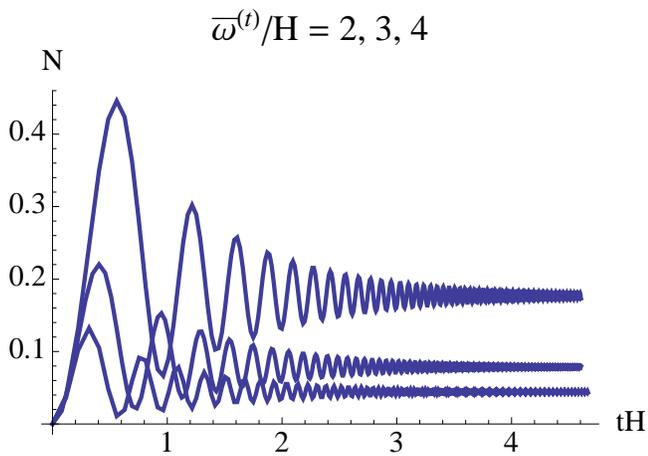}
\caption{The occupation number of produced particles, $N$, as a function of dimensionless time, $tH$, for various values of the (final) dimensionless frequency ${\bar \omega}^{(t)}/H$, in dS space. The curves are higher for lower values of ${\bar \omega}^{(t)}/H$ (higher wavelengths). At some finite time, the occupation number reaches the plateau and apart for some diminishing fluctuations remains constant at late times.}
\label{Nvs_tH}
\end{figure}

We comment here on variety of frequencies we use in this context.
A frequency certainly depends of the time evolution parameter that an observer is using.  The Schrodinger equation Eq.~(\ref{Schrod}) is written in terms of $\eta$, so $\omega$ is the mode frequency with respect to $\eta$ and not with respect to time $t$ (which we denote with $\omega^{(t)}$). From Eq.~(\ref{eta}) we see that the frequency in $t$ is $R(t)^{-3}$ times the frequency in $\eta$. Since $N$ is expressed in terms of $\bar{\omega}$, which is the frequency at the final time corresponding to $t_f$, we have
\be
  {\bar \omega}^{(t)}=e^{-3Ht_f}\bar{\omega}
  \label{omega_t}
\ee
From Eq.~(\ref{omega}) and Eq.~(\ref{omega_t}) we see that the final frequency of the harmonic oscillator states is
\be \label{master}
  {\bar \omega}^{(t)}=\omega_0e^{-Ht_f}.
\ee
This means that the frequency of the mode whose initial frequency was $\omega_0$ decreases with time, as expected. The physical frequency measured by the detector at some time $t_f$ is ${\bar \omega}^{(t)}$. Expression (\ref{master}) says that a ceratin fixed value of ${\bar \omega}^{(t)}$, at some late time, comes from a mode of much higher original frequency $\omega_0$, which was subsequently stretched by the expansion.

In Fig.~\ref{Nvs_tH} we plot the occupation number of produced particles, $N$, as a function of dimensionless time, $tH$, for different values of dimensionless frequency ${\bar \omega}^{(t)}/H$. The curves are higher for lower values of ${\bar \omega}^{(t)}/H$, which means that the occupation number in lower frequencies (higher wavelengths) is higher. At some finite time, the occupation number reaches the plateau and apart for some diminishing fluctuations remains constant at late times. This can be explained with the help of Eq.~(\ref{master}). Modes in an infinite range of frequencies are produced. The modes of any particular frequency get stretched by the expansion, but they also get constantly replaced by stretched modes whose initial frequency was smaller.

Note that ${\bar \omega}^{(t)}/H$ is the frequency of the harmonic oscillator at the final time $t_f$ at which the occupation number is evaluated. With fixed $t_f$, in order to vary ${\bar \omega}^{(t)}/H$ we need to vary $\omega_0$ since ${\bar \omega}^{(t)}=\omega_0e^{-Ht}$. Thus, different values of ${\bar \omega}^{(t)}/H$ for the fixed $t=t_f$, correspond to different $\omega_0$ which is the original frequency of the mode at the time of its creation.

\begin{figure}[htbp]
\includegraphics{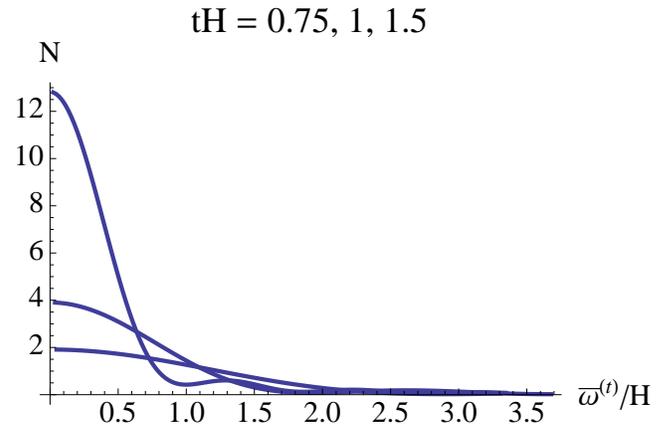}
\caption{The occupation number of produced particles, $N$, as a function of the dimensionless frequency ${\bar \omega}^{(t)}/H$, for various values of dimensionless time, $tH$, in dS space. The curves are higher (contain more particles) for later times. The plot shows the non-thermal features at very low and very high frequencies. }
\label{Nvs_w/H}
\end{figure}

In Fig.~\ref{Nvs_w/H}, we numerically evaluated the spectrum, $N$ vs. ${\bar \omega}^{(t)}/H$, of mode occupation numbers at several values of $tH$. The curves are higher for later times, which means that there are more excited modes at later times. In agreement with the plot in Fig.~\ref{Nvs_tH}, at any fixed time there are more lower frequency modes. As expected, at later times we have progressively more lower frequency (longer wavelength) modes produced by the expansion.

Comparing the distribution in Fig.~\ref{Nvs_w/H} with the thermal Planckian distribution Eq.~(\ref{N_P}) we can infer that that only the modes with frequencies comparable to the inverse Hubble radius follow the thermal distribution. Those with much lower and much higher frequencies show departure from the thermal distribution. The thermal distribution in Eq.~(\ref{N_P}) is divergent at low frequencies while  the distribution in Fig.~\ref{Nvs_w/H} is not. At high frequencies, the distribution in Fig.~\ref{Nvs_w/H} has some non-thermal oscillating features. Short wavelength modes become thermal when the expansion stretches them to the scales comparable to the Hubble radius.

In order to fit the temperature that corresponds to the particle distribution, we first express $N(\bar{\omega})$
in terms of time $t$. Then we plot $\ln(1+1/N)$ versus ${\bar \omega}^{(t)}/H$ and find the slope $\beta$. Finally, we correct for the factor in Eq.~(\ref{omega_t}) which implies
\be
  T=e^{-3Ht_f}\beta^{-1}(t_f).
  \label{temp}
\ee
This is then the temperature measured at some moment $t_f$, seen by the observer who is using time $t$ as an evolution parameter.

\begin{figure}[htbp]
\includegraphics{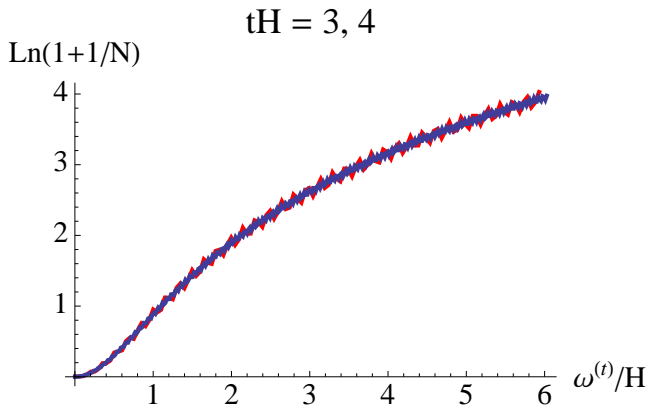}
\caption{Here we plot $\ln(1+1/N)$ versus ${\bar \omega}^{(t)}/H$ for the values of $tH=3,4$ for the dS space. The curves for $tH=3$ (black) and $tH=4$ (red) are almost indistinguishable which indicates that the temperature does not change with time. Since only the modes with ${\bar \omega}^{(t)}/H \sim 1$ are thermal, we fit the temperature only in that segment of the plot and find it to be proportional to the Hubble constant. Very high and very low frequencies are not thermal.}
\label{LnN_versus_w}
\end{figure}

In Fig.~\ref{LnN_versus_w} we plot $\ln(1+1/N)$ versus ${\bar \omega}^{(t)}/H$ for various values of $tH$. Here we see that the curves corresponding to $tH=3$ (blue) and $tH=4$ (red) are almost indistinguishable. Since $\ln(1+1/N)$ versus ${\bar \omega}^{(t)}/H$ will give us information about the temperature, this means that the temperature does not change at late times.

Let us analyze the plot in Fig.~\ref{LnN_versus_w}. We mentioned that only the modes with frequencies comparable to the Hubble constant (inverse Hubble radius) follow the thermal distribution. Therefore, in order to get the temperature of the produced radiation, we fit the straight line only in the segment of the plot corresponding to the neighborhood of ${\bar \omega}^{(t)}/H =1$. From Fig.~\ref{Nvs_w/H} we see that most of the modes belong to that range (say we sum up the number of modes within one order of magnitude, from  ${\bar \omega}^{(t)}/H = 0.15$ to ${\bar \omega}^{(t)}/H = 1.5$). The number of modes of very high frequency is very low since $N(\bar{\omega}^{(t)})$ drops in that regime quickly, while the number of modes with very low frequency is very low since that regime is a very short segment on the  $N(\bar{\omega}^{(t)})$ curve. Thus fitting the curve in the omega ${\bar \omega}^{(t)}/H \sim 1$ segment would give a pretty realistic result for the temperature of the produced radiation since most of the produced particles will be in that regime.
Applying Eq.~(\ref{temp}), the fit gives for the slope $\beta \sim 1$ in units of $1/H$ which implies
\be
T \sim H  .
\ee
As expected, the temperature of the produced radiation is proportional to the Hubble constant, which is in turn proportional to the value of the cosmological constant $\lambda$.

The modes with much lower and much higher frequencies than the Hubble constant do not follow the thermal distribution. However, at some later time, modes that were created with short wavelengths become thermal when the expansion stretches them to the scales comparable to the Hubble radius.

\section{Particle production in anti-de Sitter space-time}

Now we apply the analogous study to AdS space with the negative cosmological constant, i.e. $\lambda < 0$.
For this purpose, we set $k=-1$ in Eq.~(\ref{metric}). In that case, the solution for the scale factor is given by $R(t)=H^{-1}\cos(Ht)$ where $H$ is $H =\sqrt{|\lambda |/3}$. Therefore the time dependent AdS metric takes on the same form as dS metric given in Eq.~(\ref{de_sit}), apart from the different scale factor $R(t)$.
For the decaying segment of $\cos(Ht)$, this time dependent solution describes a collapsing universe under the influence of the negative constant vacuum  energy density represented by the negative cosmological constant.
The matrices ${\bf M}$ and ${\bf N}$ are now given by Eq.~(\ref{de M}) and Eq.~(\ref{de N}), with the substitution of $k=-1$
\bea
  {\bf M}_{kk'}&=&4\pi\int drr^2\frac{1}{\sqrt{1+r^2}}f_kf_{k'},\label{adM}\\
  {\bf N}_{kk'}&=&4\pi\int drr^2\sqrt{1+r^2}f'_kf'_{k'}.
  \label{adN}
\eea

\begin{figure}[htbp]
\includegraphics{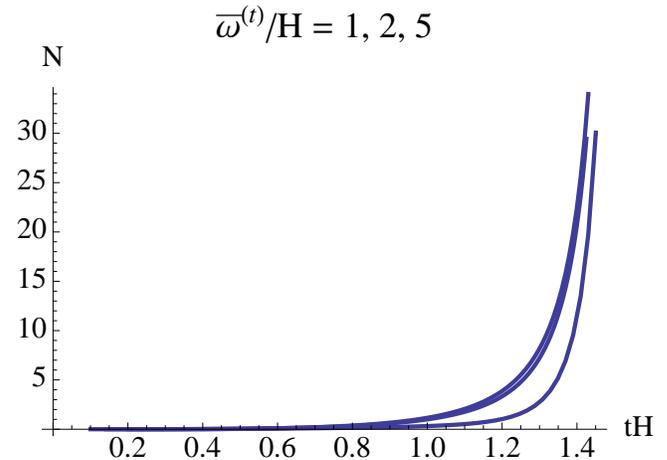}
\caption{The occupation number of produced particles, $N$, as a function of dimensionless time, $tH$, for various values of dimensionless frequency ${\bar \omega}^{(t)}/H$, in AdS space. The curves are higher for higher values of ${\bar \omega}^{(t)}/H$.  The occupation number diverges in finite time, indicating the Big Crunch. }
\label{aN_versus_tH}
\end{figure}

The occupation number is given again by Eq.~(\ref{Occ_Num}).
Repeating the procedure from the previous section, we plot $N$ versus $tH$ for various values of ${\bar \omega}^{(t)}/H$ in Fig.~\ref{aN_versus_tH}. The Figure shows that for different values of dimensionless frequency ${\bar \omega}^{(t)}/H$, the occupation number increases in all frequency modes. However, the curves are higher for higher values of ${\bar \omega}^{(t)}/H$, which means that the occupation number in higher frequencies (lower wavelengths) increases faster at later times. In contrast with dS space where the occupation number at some fixed final frequency is constant at late times, the occupation number in AdS space diverges, and it does this in finite time, indicating the big crunch.

\begin{figure}[hbp]
\includegraphics{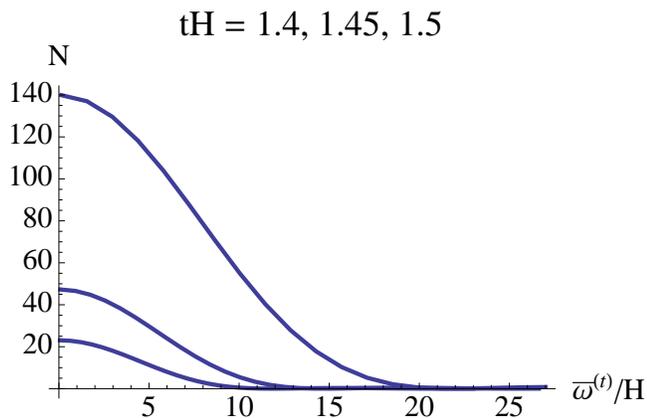}
\caption{The occupation number of produced particles, $N$, as a function of dimensionless frequency ${\bar \omega}^{(t)}/H$, for various values of dimensionless time, $tH$, in AdS space. The occupation number at all frequencies grows as $tH$ increases, however the high frequency modes grow at a higher rate.}
\label{aN_versus_w}
\end{figure}

In Fig.~\ref{aN_versus_w}, we have numerically evaluated the spectrum, $N$ versus ${\bar \omega}^{(t)}/H$, for several values of $tH$. The curves are higher for later times, which means that there are more excited modes at later times. As expected, at later times we have progressively more higher frequency modes since the space is collapsing.

We may now try to fit the temperature that corresponds to the particle distribution. We first express $N(\bar{\omega})$
in terms of time t and then plot $\ln(1+1/N)$ versus ${\bar \omega}^{(t)}/H$ for various values of $tH$.  We could then fit the straight line through the curves and find the slope $\beta$ for each of them. Eq.~(\ref{eta}) tells us that the frequency in $t$ is $R(t)^3$ times the frequency in $\eta$, and at time $t_f$, this implies
\be
  {\bar \omega}^{(t)}=\frac{H^3}{\cos^3(Ht)}\bar{\omega}
  \label{anti_omega_t}
\ee
 Therefore the temperature seen by the observer whose parameter of evolution is t should be
\be
  T=\frac{H^3}{\cos^3(Ht)}\beta^{-1}(t_f).
  \label{anti_temp}
\ee

\begin{figure}[htbp]
\includegraphics{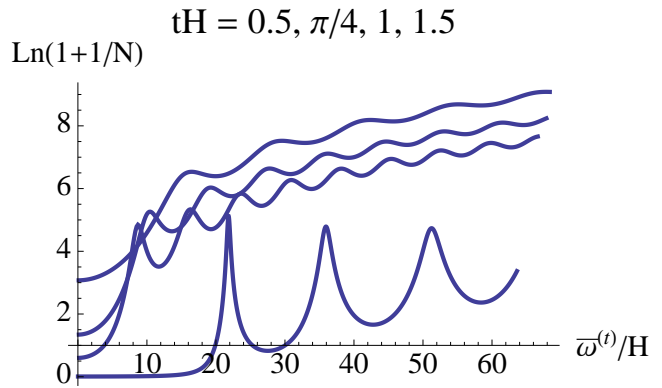}
\caption{Here we plot $\ln(1+1/N)$ versus ${\bar \omega}^{(t)}/H$ for various values of $tH$ for the AdS space. The curves are lower for later times. The slope of the best fit straight line is supposed to give the temperature of radiation. However, the plot shows that the oscillations become greater as $tH$ increases, making the temperature less and less defined toward the Big Crunch.}
\label{aLnN_versus_w}
\end{figure}

However, following this procedure we find it very difficult to precisely fit the temperature. Nominally, from Eq.~(\ref{anti_temp}), the temperature diverges as $tH$ approaches the time of the Big Crunch. However, from Fig.~\ref{aLnN_versus_w} we see that as $tH$ increases the oscillations of the curves increase making the temperature less and less defined toward the Big Crunch.
This is in agreement with the fact that AdS space (unlike dS space) does not have an event horizon which could cause thermal radiation.

\section{Conclusion}

In this paper, we study the time-dependent evolution of the scalar field propagating in the backgrounds of an expanding dS and contracting AdS universe. Using the functional Schrodinger formalism we are able to probe the time dependent regime which is out of the reach of the standard approximations like the Bogolyubov method. In both cases, the evolution of fluctuations is governed by the harmonic oscillator equation with time dependent frequency. In general, the harmonic oscillator equation governs the evolution of quantum fluctuations and particle production. We solved the equations of motion and found the time-dependent wave function for the system. The wave function overlap between the initial state (which is the vacuum) and the final state at some arbitrary later time gave us the time-dependent occupation number of produced particles.

In the case of an expanding dS universe we explicitly showed that the frequency of fluctuations produced at a certain moment diminish in time, while the distribution of the created particles quickly approaches the thermal radiation of the dS space.  In the case of a contracting AdS universe we showed that the frequency of fluctuations produced at a certain moment grow in time and the temperature of radiation diverges as the big crunch is approaching, which is in agreement with earlier studies. We also explicitly demonstrated that when a mode of a certain frequency is created in dS space it does not automatically follow a thermal distribution with the temperature proportional to the value of the Hubble parameter. Modes whose frequencies are much larger or much smaller than the Hubble parameter do not follow a thermal distribution, while those of the order of the Hubble parameter do. Very high frequency modes are not thermal since they were produced from a non-thermal initial vacuum state and it takes some time for them to stretch and thermalize. Very low frequency modes which are larger than a horizon are frozen, and therefore non-thermal. As the high frequency modes get stretched by expansion, and their frequency becomes comparable to the Hubble parameter, they become thermal. In AdS case, towards the Big Crunch, though the temperature nominally diverges, fluctuations of the spectrum grow and the temperature becomes less and less defined.

The functional Schrodinger formalism we used here is different from the standard techniques used so far in similar studies. In \cite{Sekiwa:2008gk}, some objections to the standard quantum tunneling treatment (see for example \cite{Medved:2002zj,Spradlin:2001pw,Nakayama:1988wy}) are brought up, while in \cite{Volovik:2008ww}
it was suggested that the dS space is stable against the Hawking radiation. Our independent approach supports the standard picture of thermal radiation in dS space.

\begin{acknowledgments}
 D.S. acknowledges the financial support from NSF, grant number PHY-0914893. E.G. was supported by NASA ATP grant NNX07AG89G to Case Western Reserve University.
\end{acknowledgments}

\appendix

\section{Number of particles radiated as a function of time}
\label{Number}

We use the simple harmonic oscillator basis states but at a frequency $\bar{\omega}$ to keep track of the different $\omega$'s in the calculation. To evaluate the occupation numbers at time $t>t_f$, we need only set $\bar{\omega}=\omega(t_f)$. So
\be
  \phi_n(b)=\left(\frac{m\bar{\omega}}{\pi}\right)^{1/4}\frac{e^{-m\bar{\omega}b^2/2}}{\sqrt{2^nn!}}{\cal{H}}_n(\sqrt{m\bar{\omega}}b)
\ee
where ${\cal{H}}_n$ are the Hermite polynomials. Then Eq.~(\ref{c_n}) together with Eq.~(\ref{psi}) gives
\bea
  c_n&=&\left(\frac{1}{\pi^2\bar{\omega}\rho^2}\right)^{1/4}\frac{e^{i\alpha}}{\sqrt{2^nn!}}\int d\xi e^{-P\chi^2/2}{\cal{H}}_n(\xi)\nonumber\\
    &=&\left(\frac{1}{\pi^2\bar{\omega}\rho^2}\right)^{1/4}\frac{e^{i\alpha}}{\sqrt{2^nn!}}I_n,
\eea
where
\be
  P\equiv1-\frac{i}{\bar{\omega}}\left(\frac{\rho_{\eta}}{\rho}+\frac{i}{\rho^2}\right).
\ee

To find $I_n$ consider the corresponding integral over the generating function for the Hermite polynomials
\bea
  J(z)&=&\int d\xi e^{-P\xi^2/2}e^{-z^2+2z\xi}\nonumber\\
    &=&\sqrt{\frac{2\pi}{P}}e^{-z^2(1-2/P)}.
\eea
Since
\be
  e^{-z^2+2z\chi}=\sum_{n=0}^{\infty}\frac{z^n}{n!}{\cal{H}}_n(\xi)
\ee
\be
  \int d\xi e^{-P\xi^2/2}{\cal{H}}_n(\chi)=\frac{d^n}{dz^n}J(z)\Big{|}_{z=0}.
\ee
Therefore
\be
  I_n=\sqrt{\frac{2\pi}{P}}\left(1-\frac{2}{P}\right)^{n/2}{\cal{H}}_n(0).
\ee
Since
\be
  {\cal{H}}_n(0)=(-1)^{n/2}\sqrt{2^nn!}\frac{(n-1)!!}{\sqrt{n!}}, \hspace{2mm}\text{$n$=even}
\ee
and ${\cal{H}}_n(0)=0$ for odd $n$, we find the coefficients $c_n$ for even values of $n$,
\be
  c_n=\frac{(-1)^{n/2}e^{i\alpha}}{(\bar{\omega}\rho^2)^{1/4}}\sqrt{\frac{2}{P}}\left(1-\frac{2}{P}\right)^{n/2}\frac{(n-1)!!}{\sqrt{n!}}.
\ee
For odd $n$, $c_n=0$.

We next find the number of particles produced. Let
\be
  \chi=\left|1-\frac{2}{P}\right|.
\ee
Then
\bea
  N(t,\bar{\omega})&=&\sum_{n=even}n|c_n|^2\nonumber\\
      &=&\frac{2}{\sqrt{\bar{\omega}}|P|}\chi\frac{d}{d\chi}\sum_{n=even}\frac{(n-1)!!}{n!!}\chi^n\nonumber\\
      &=&\frac{2}{\sqrt{\bar{\omega}}|P|}\chi\frac{d}{d\chi}\frac{1}{\sqrt{1-\chi^2}}\nonumber\\
      &=&\frac{2}{\sqrt{\bar{\omega}}|P|}\frac{\chi^2}{(1-\chi^2)^{3/2}}.
\eea
Inserting the expressions for $\chi$ and $P$, leads to
\be
  N(t,\bar{\omega})=\frac{\bar{\omega}\rho^2}{\sqrt{2}}\left[\left(1-\frac{1}{\bar{\omega}\rho^2}\right)^2+\left(\frac{\rho_{\eta}}{\bar{\omega}\rho}\right)^2\right].
\ee

In summary, we have found the occupation number of modes as a function of $\rho$ which is a function of time as given by the non-linear differential equation Eq.~(\ref{rho}).

\end{document}